\documentclass{article}
\usepackage{multirow,booktabs,makecell,spconf,amsmath,graphicx,amsfonts,color,subfigure}
\usepackage[numbers,sort&compress]{natbib}

\title{MIMO-DBnet: Multi-channel Input and Multiple Outputs DOA-aware Beamforming Network for Speech Separation}
\name{Yanjie Fu$^{1}$, Haoran Yin$^{1}$, Meng Ge$^{1,2}$, Longbiao Wang$^{1}$, Gaoyan Zhang$^{1}$, Jianwu Dang$^{1}$, \\ Chengyun Deng$^{3}$, Fei Wang$^{3}$}
\address{$^{1}$College of Intelligence and Computing, Tianjin University, Tianjin, China\\
  $^2$The Chinese University of Hong Kong, Shenzhen, China \\
  $^{3}$Beijing Xiaoju Technology Co., Ltd., Beijing, China\\}
\email{$\{$fuyanjie, haoran\_yin, gemeng, longbiao\_wang$\}$@tju.edu.cn}
\begin{document}
\ninept
\maketitle
\begin{abstract}
Recently, many deep learning based beamformers have been proposed for multi-channel speech separation. Nevertheless, most of them rely on extra cues known in advance, such as speaker feature, face image or directional information. In this paper, we propose an end-to-end beamforming network for direction guided speech separation given merely the mixture signal, namely MIMO-DBnet. Specifically, we design a multi-channel input and multiple outputs architecture to predict the direction-of-arrival based embeddings and beamforming weights for each source. The precisely estimated directional embedding provides quite effective spatial discrimination guidance for the neural beamformer to offset the effect of \textit{phase wrapping}, thus allowing more accurate reconstruction of two sources' speech signals. Experiments show that our proposed MIMO-DBnet not only achieves a comprehensive decent improvement compared to baseline systems, but also maintain the performance on high frequency bands when \textit{phase wrapping} occurs.
\end{abstract}
\begin{keywords}
Speech separation, direction-of-arrival estimation, phase wrapping, spatial discrimination, MVDR beamformer
\end{keywords}
\section{Introduction}
\label{sec:intro}
Speech separation aims to extract all individual speech signals from the observed mixture speech, which is a high-demand front-end technology for various real-world speech applications, such as speech recognition \cite{higuchi2017online, chen2018multi} and speaker verification \cite{rao2019target, xu2021target}. 

For quite some time, multi-channel speech separation solutions have attracted much research attention due to the benefit of spatial information in the microphone array setup. Various inter-channel features including, interaural phase difference (IPD), interaural time difference (ITD), interaural level difference (ILD) and normalized cross-correlation (NCC), used to be considered as the most reliable spatial features. However, when the spacing between elements of microphone array is not small enough to meet the spatial Nyquist criterion, the \textit{spatial aliasing} occurs \cite{Johnson1993ArraySP}. This leads to \textit{phase wrapping} on high frequencies, i.e., the phase difference becomes ambiguous.
For a two-element microphone array with a spacing of 4.25cm, the spatial aliasing frequency is 4 kHz \cite{dmochowski2008spatial}, i.e., \textit{phase wrapping} occurs above 4 kHz, which means the inter-channel features (differences or correlations) become less discriminative and cannot be utilized to accurately distinguish one source from another in terms of their spatial information on high frequencies.

Recently, data-driven neural network methods have brought remarkable progress in multi-channel speech separation literature by leveraging neural network's powerful nonlinear mapping ability. For example, the deep learning-based beamformers are well studied to estimate speech and noise covariance matrices and beamforming weights in an end-to-end fashion. Most of them introduce extra cues to make up the less discriminative spatial feature when \textit{phase wrapping} issue occurs, including speaker cue \cite{9746221} from an enrolled reference utterance, visual cue \cite{gu2020multi} from face tracking and lip movements, and direction cue \cite{xu20_interspeech, zhang2021ADL, xu2021GRNN} from an extra visual detection system.


In blind source separation scenarios, the reference signal as well as other cues are not available and only the observed mixture signal can be leveraged to process. We argue that it is still possible to compensate the absence of spatial discrimination due to \textit{phase wrapping}, even with the only received mixture input.
In this work, to improve the spatial discrimination and relieve the performance decline problem under \textit{phase wrapping} conditions with significant reverberation, a multi-channel input and multiple outputs DOA-aware purely neural beamformer, termed as MIMO-DBnet, is proposed to explicitly guide speech separation task with segregated directional information, even on small angular spacing cases. The only input is the multi-channel mixture audio, i.e., there is no requirement for providing any extra auxiliary information in our system. Specifically, the speech and interference covariance matrices for each source are obtained first and two parallel branches then estimate the DOAs of each source and their corresponding speech signals simultaneously. By respectively perform DOA estimation for each individual speaker rather than for all sources in the mixture once, ambiguous phase differences are turned into discriminative directional embeddings and the DOAs are precisely predicted. The obtained DOA-based directional embeddings further provide spatial discrimination to fill the blank left by inter-channel features and help the network to accurately segregate the speech signals of two speakers, even on high frequencies effected by \textit{phase wrapping} issue. Experimental results show that both speech separation performance and speech intelligibility are substantially improved compared to existing state-of-the-art methods.
\begin{figure*}[ht]
\centering
	\includegraphics[width=1\linewidth]{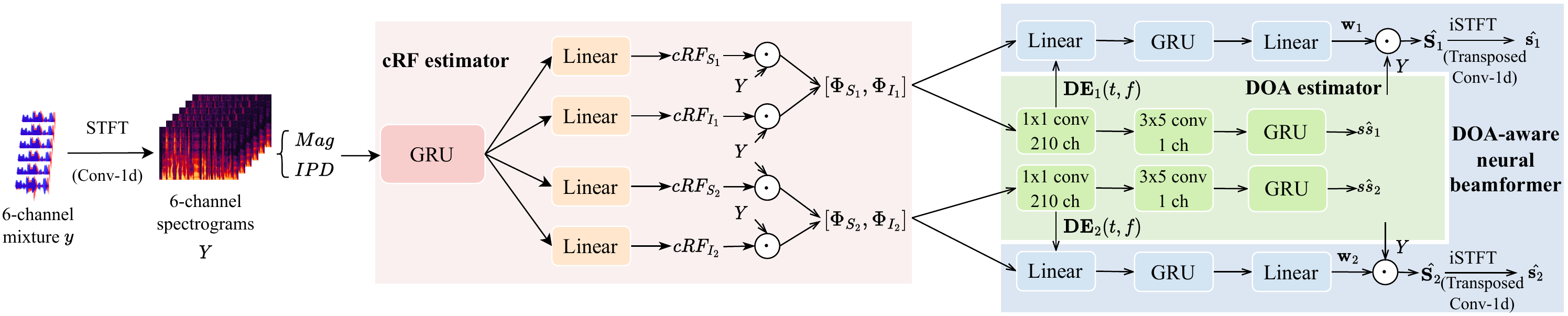}
	\vspace{-20px}
	\caption{Illustration of the architecture of our proposed MIMO-DBnet. cRF estimator produces cRFs to obtain speech and covariance matrices for both speakers as input of following modules. DOA estimator is introduced for explicit spatial discrimination guidance to neural beamformer with the predicted directional embedding $\mathbf{DE}_i(t, f)$. DOA-aware neural beamformer predicts the beamforming weights for each source. All components in both the time and frequency domain of our framework are trainable, which allows backpropagation all around.}
	\label{fig:framework}
	\vspace{-10px}
\end{figure*}
\vspace{-5px}
\section{Generalized RNN-beamformer baseline}
\label{sec:model}

Consider the input mixture waveform $\mathbf{y} \in \mathbb{R}^{L\times M}$, where $L$ is denoted as the number of audio samples and $M$ as the number of audio channels, the corresponding time-frequency representations after STFT can be formulated as
\begin{equation}
Y(t,f)=S(t,f)+I(t,f)
\end{equation}
where $S$ stands for the target speech and $I$ represents the sum of interfering speakers' speech and noise (if any).

GRNN-BF proposed in \cite{xu2021GRNN} contains two modules, complex-valued ratio filter (cRF) estimator and RNN beamformer.
The cRFs are predicted by cRF estimator and then used to calculate the target speech and interference covariance matrices. The RNN beamformer accepts the target speech and interference covariance matrices and generates frame-level beamforming weights for the target speaker.

For each T-F (time-frequency) unit of mixture, GRNN-BF applies the estimated cRF to its neighbouring units following
\begin{equation}
\begin{split}
\hat{\mathbf{S}}_i(t, f)=\sum_{\tau_{1}=-K}^{\tau_{1}=K}\sum_{\tau_{2}=-K}^{\tau_{2}=K}\mathrm{cRF}_{\mathbf{S}_i}\left(t,f,\tau_{1},\tau_{2}\right)\\
*\mathbf{Y}\left(t+\tau_{1},f+\tau_{2}\right)
\end{split}
\label{eq:speech}
\end{equation}
where $\hat{\mathbf{S}}_i(t, f)$ indicates the $i$-th speaker's speech estimated by ${cRF}_{\mathbf{S}_i}$, and $K$ defines the number of neighboring T-F bins. Likewise, the $i$-th speaker's interference noise $\hat{\mathbf{I}}_i(t, f)$ can be obtained by ${cRF}_{\mathbf{I}_i}$ in a similar manner. Then, the frame-wise $i$-th speaker's speech covariance matrix can be calculated with the estimated target speech and its conjugate transpose by a layer normalization \cite{ba2016layer} shown in Eq. (\ref{eq:covmat}). Another layer normalization is applied for computing speaker $i$'s corresponding interference noise covariance matrix $\mathbf{\Phi}_{\mathbf{I}_i}(t, f)$.
\begin{equation}
\mathbf{\Phi}_{\mathbf{S}_i}(t, f) = \text{LayerNorm}(\hat{\mathbf{S}}_{i}(t, f) \hat{\mathbf{S}}_{i}^{\mathrm{H}}(t, f))
\label{eq:covmat}
\end{equation}
GRNN-BF believes that a better beamformer solution can be directly learned from the speech and noise covariance matrices by neural network. Thus only one unified RNN model is applied to predict the frame-level beamforming weights directly from the covariance matrices, which can be formulated as
\begin{equation}
\mathbf{w}_{i}(t, f) =\mathbf{R N N}\left(\left[\boldsymbol{\Phi}_{\mathbf{S}_i}(t, f), \mathbf{\Phi}_{\mathbf{I}_i}(t, f)\right]\right)
\label{eq:rnnbf}
\end{equation}
Here, $[\cdot,\cdot]$ denotes the concatenation operation of the $i$-th speaker's speech and interference noise covariance matrices. Next, the STFT representation of mixture $Y(t,f)$ is beamformed with the estimated beamforming weights $\mathbf{w}_{i}(t, f)$ to obtain the estimation of $i$-th target speaker's speech spectrogram $\hat{\mathbf{S}_i}(t,f)$, which can be formulated as
\begin{equation}
\hat{\mathbf{S}_i}(t, f) =\mathbf{w}_{i}^{\mathrm{H}}(t,f) \mathbf{Y}(t, f)
\label{eq:beamformed}
\end{equation}
Finally, the separated time-domain waveform of the $i$-th speaker's speech $\hat{\mathbf{s}_i}$ can be converted from the beamformed spectrum $\hat{\mathbf{S}_i}(t, f)$ using iSTFT. All of the covariance matrices and beamforming weights are complex-valued and the real and imaginary parts of these complex-valued tensors are concatenated from start to end.

\section{MIMO-DBnet Architecture}
GRNN-BF \cite{xu2021GRNN} requires prior known DOA of target speaker, which is calculated according to the location of target speaker in the video view captured by a $180^{\circ}$ wide-angle camera, thus does not suit applications for blind source separation where only observed mixture signal is available. Meanwhile, the DOA is roughly estimated and its DOA estimator cannot be jointly optimized with the neural beamformer, which limits the upper bound performance of the system.

In this section, we modify GRNN-BF to alleviate the spatial ambiguity on high frequency bands caused by \textit{phase wrapping} issue \cite{8682470} with the assitance of explicit guidance from discriminative directional information merely using the given mixture signal.
\subsection{Overview of MIMO-DBnet}
We refer our proposed framework to multi-channel input and multiple outputs DOA-aware beamforming network (MIMO-DBnet). Fig. \ref{fig:framework} shows the detailed architecture of our proposed MIMO-DBnet, which consists of cRF estimator, DOA estimator and DOA-aware neural beamformer. Specifically, the magnitude of reference channel and interaural phase difference (IPD) features of five microphone pairs are extracted from the STFT representation converted from multi-channel mixture audio signal. Unlike GRNN-BF, the directional feature captured by camera is not available here and we pass magnitude rather than log power spectrum (LPS) along with IPD to the cRF estimator. We replace the dialated Conv-1D blocks with a GRU module in cRF estimator due to consideration of saving GPU memory usage. After obtaining the cRFs for each speaker's speech and interference, multi-channel target speech and corresponding interference can be computed with Eq. (\ref{eq:speech}). Then, the speech and interference covariance matrices are calculated with two layer normalization \cite{ba2016layer} modules shown in Eq. (\ref{eq:covmat}). The next step is to predict beamforming weights from the covariance matrices. In this work, we propose to introduce a tiny DOA estimator to offer direction discrimination for the GRU-based beamformer, even on high frequencies where \textit{phase wrapping} occurs, to learn more accurate beamforming weights and exhibit more effective speech separation capability.
\vspace{-5px}
\subsection{DOA estimator}
For each source, the DOA estimator accepts its speech and interference covariance matrices and output the frame-level spatial spectrum related to its DOA. The likelihood-based spatial spectrum coding \cite{he2018deep} is based on the assumption that the probability of source being at each individual angle follows Gaussian distribution that maximizes at the ground truth DOA. Formulaically, we adopt a 210-dimensional vector $ss(\theta)$ for each time frame to include the probabilities of sound source locating at 210 individual directions (the azimuth $\theta$ ranges from $-15^{\circ}$ to $195^{\circ}$). We assign the values of vector $ss(\theta)$ as below:
\begin{equation}
ss(\theta)=e^{-{d\left(\theta,\theta^{\prime}\right)^{2}}/{\sigma^{2}}}
\end{equation}
where $\theta^{\prime}$ is the ground truth DOA of one source, $\sigma$ is a predefined constant that controls the width of the Gaussian function and $d(\cdot,\cdot)$ denotes the angular distance.

The DOA estimator is composed of two convolutional layers and a GRU module. The first convolutional layer projects the covariance matrices into the DOA space to obtain 210-dimensional directional embedding for each T-F unit by expanding feature dimension to the number of directions. The second one convolves along the time and DOA axes and aggregates features across all frequency bins to generate initial spatial spectrums for each time frame. The following GRU module learns the temporal context information among all time frames to polish the frame-level spatial spectrum before outputing it. We select the mean squared error to measure the disparity between the estimated and the ideal frame-level spatial spectrum for the $i$-th speaker $\hat{ss}_i$ and $ss_i$, which can be represented as
\begin{equation}
\mathcal{L}_{\text{SS}_i}=||\hat{ss_i}-ss_i||^2
\end{equation}
\subsection{DOA-aware neural beamformer}
To address the problem of phase difference ambiguity on high frequencies when using inter-channel features, we additionally pass the DOA estimator's intermediate output, the directional embedding, that indicates from which azimuth direction the speech comes in each T-F bin to our proposed DOA-aware neural beamformer.

Technically speaking, the estimated directional embeddings for each speaker $\{\mathbf{DE}_i(t, f)\}_{i=1}^2$ are concatenated with the speech and interference covariance matrices of corresponding speaker along feature dimension and respectively fed into two parallel estimators to generate the beamforming weights for each speaker following Eq. (\ref{eq:dbrnnbf}) and thus can more accurately separate the two speaker's speech.
\begin{equation}
\mathbf{w}_{i}(t, f) =\mathbf{R N N}\left(\left[\boldsymbol{\Phi}_{\mathbf{S}_i}(t, f), \mathbf{\Phi}_{\mathbf{I}_i}(t, f), \mathbf{DE}_i(t, f)\right]\right)
\label{eq:dbrnnbf}
\end{equation}

In this way, the DOA estimator explicitly performs an auxiliary task for blind source separation. The introduced directional embedding just complements the ambiguity exists in IPDs on high frequency bands, which enables the neural beamformer to regain spatial discrimination even if \textit{phase wrapping} issue occurs.
For end-to-end training, the time-domain weighted source-to-distortion ratio (wSDR) loss proposed in \cite{choi2018phase} is chosen as the speech separation criterion, which can be computed as follows:
\begin{equation}
\left\{\begin{array}{l}
\mathcal{L}_{\text{SDR}}(\mathbf{x}, \hat{\mathbf{x}})=\frac{\langle \mathbf{x}, \hat{\mathbf{x}}\rangle}{\|\mathbf{x}\|\|\hat{\mathbf{x}}\|} \\
\vspace{1pt}
\alpha=\frac{\|\mathbf{s}_i\|^2}{\|\mathbf{s}_i\|^2+\|\mathbf{i}_i\|^2} \\
\vspace{1pt}
\mathcal{L}_{\text{wSDR}_i}(\mathbf{y}, \mathbf{s}_i, \hat{\mathbf{s}}_i)\\=-\alpha \mathcal{L}_{\text{SDR}}(\mathbf{s}_i, \hat{\mathbf{s}}_i)-
(1-\alpha) \mathcal{L}_{\text{SDR}}(\mathbf{i}_i, \hat{\mathbf{i}}_i)
\end{array}\right.
\end{equation}
where $\mathbf{y}$, $\mathbf{s}_i$ and $\hat{\mathbf{s}}_i$ represent mixture, target speech and estimated target speech signal of the $i$-th speaker respectively, $\mathbf{i}_i=\mathbf{y}-\mathbf{s}_i$ indicates its corresponding ground-truth interference signal and $\hat{\mathbf{i}}_i=\mathbf{y}-\hat{\mathbf{s}}_i$ stands for the estimated interference speech signal. The wSDR losses of both speakers are added up as the speech separation loss.

For optimizing the whole MIMO-DBnet system, we use a multi-task loss as the training objective:
\vspace{-5pt}
\begin{equation}
\mathcal{L}_{\text{MIMO-DBnet}} = \alpha * \sum_{i=1}^2 \mathcal{L}_{\text{SS}_i} + \beta * \sum_{i=1}^2 \mathcal{L}_{\text{wSDR}_i}
\end{equation}
where $\alpha$ and $\beta$ are the hyper parameters that control the weights of DOA estimation task and speech separation task, respectively.
\subsection{Angle sorting training strategy}
We design angle sorting training strategy, sorting all sources' labels in ascending order by DOAs. In this way, permutations of DOAs and speech signals are not mismatched and MIMO-DBnet can be clear about with which branch to estimate which source during training. 

\vspace{-2px}
\section{Experiments and results}
\label{sec:exps}
\vspace{-3px}
\subsection{Dataset}
We simulate 6-channel reverberant speech from VCTK corpus \cite{yamagishi2019cstr} with a non-uniform linear microphone array with spacings of 0.04 m, 0.04 m, 0.12 m, 0.04 m, 0.04 m. We use pyroomacoustics \cite{scheibler2018pyroomacoustics} to randomly simulate RIRs of 50 different rooms for every dataset split. The sizes (length, width, height) of the rooms are ranged from 4m, 3m, 3m to 15m, 15m, 3.5m. The RT60s are sampled between 0.2s and 0.7s. More details about the simulated dataset can be found in \cite{yin22b_interspeech} and https://github.com/TJU-haoran/VCTK-16k-simulated. After the simulation, we randomly select two source signals of different speakers from the same room and scale the source signals according to a random signal-to-interference ratio (SIR) between -10 dB and 10 dB before mixing. Our dataset contains 40,000 utterances (44 hours, 90 speakers) for training and 1000 utterances (1.1 hours, 10 speakers) each for validation and testing. 
\vspace{-3px}
\subsection{Experimental setup}
We train all the systems in a segment-wise mode with 4 seconds long audio segments. We use one warm-up epoch and Adam optimizer \cite{kingma2014adam}. The initial learning rate is set to 10e-4 and the max norm of the gradients is set to 3. For the first 5 epochs, the MIMO-DBnet is trained with emphasis on DOA estimation ($\alpha$ = 5, $\beta$ = 1). And for rest epochs, $\alpha$ and $\beta$ are set to 1 and 10, respectively. For ablation study, we train MIMO-DBnet w/o DE and MIMO-DBnet w/o DE\&IPD with only separation loss after removing the DOA estimator of MIMO-DBnet. We apply uPIT strategy \cite{kolbaek2017multitalker} when training all systems except MIMO-DBnet which is trained with angle sorting training strategy. The model achieves the best performance on the validation set is chosen after 30 epochs. STFT is conducted with 512-point FFT along 32ms Hamming window with 50\% stride, and the output feature dimension is 257. The STFT and iSTFT are implemented with fixed convolutional encoder and decoder \cite{gu2020multi}. The cosIPDs \cite{wang2018multi} are computed from five pairs between the first microphone and the rest microphones. For MIMO-DBnet and MIMO-DBnet w/o DE, the 1st channel's magnitude together with cosIPDs are taken as input. MIMO-DBnet w/o DE\&IPD is input with magnitude of 6-channel spectrograms instead.
The cRF estimator is a 2-layer uni-directional GRU with 500 hidden units followed by 4 FC layers with ReLU activation \cite{glorot2011deep}. The GRU module in DOA estimator is a 2-layer uni-directional GRU with 210 hidden units and every neural beamformer contains a FC layer and a 2-layer uni-directional GRU with 300 hidden units followed by a FC layer. The size of cRF $K\times K$ is set to $3\times 3$. The constant $\sigma$ in spatial spectrum coding is set to 8. We conduct all our experiments with a batch size of 4.

\vspace{-2px}
\subsection{Results and discussion}
We evaluate the 2-speaker separation performance of different sytems in terms of SI-SDR, PESQ and WER\footnote{\label{nemo}We use pretrained Conformer-CTC Large to evaluate the ASR performance. See https://huggingface.co/nvidia/stt\_en\_conformer\_ctc\_large.} on VCTK-16k-simulated. The reverberant speech of each source is taken as reference signal to measure the error of separated speech. The DOA estimation performance is also tested with accuracy ($5^{\circ}$ tolerance) and mean absolute error if there exists DOA estimator. The angle sorting training strategy is only applied in training stage as angle order is not given when inferencing and all results are the ones computed from the permutation with the highest SI-SDR score.
\vspace{-5px}
\subsubsection{Overall speech separation performance}
We first compare our proposed MIMO-DBnet to FaSNet with TAC with NCC feature (which is referred to as FaSNet+TAC below) proposed in \cite{9054177}. As shown in Table 1, performance of our proposed MIMO-DBnet obtains 1.96 dB absolute improvement on SI-SDR compared to FasNet+TAC. In metric of PESQ, it can be seen that MIMO-DBnet and MIMO-DBnet w/o DE achieve \textbf{32.4\%} and 24.9\% relative improvements compared to FasNet+TAC (i.e., 2.29 vs. 1.73 and 2.16 vs. 1.73), which demonstrates that the frequency domain neural beamformer is more helpful for restoring speech quality and thus can benefit downstream tasks such as speech recognition.
As shown in Table \ref{tbl:wer}, the WER results of our proposed MIMO-DBnet and MIMO-DBnet w/o DE achieves \textbf{35.4\%} and 31.8\% absolute improvements compared with that of FasNet+TAC, respectively.

\begin{table}[t]
\centering
\fontsize{9}{8}\selectfont
\caption{Objective measurement (SI-SDR and PESQ) results of baselines and our proposed MIMO-DBnet on the entire test data and among different and same gender mixtures.}
\vspace{5px}
    \setlength{\tabcolsep}{1mm}{
        \begin{tabular}{l c c c c c c}
        \toprule
        \multirow{2}{*}{\textbf{Systems}\ \textbf{/\ Metrics}} & \multicolumn{3}{c}{\textbf{SI-SDR $\uparrow$}} & \multicolumn{3}{c}{\textbf{PESQ $\uparrow$}}  \\
        \cmidrule(l){2-4} \cmidrule(l){5-7} 
        & Diff. & Same & Avg. & Diff. & Same & Avg.
        \\
		\midrule
		Mixture & -0.00 & 0.01 & 0.00 & 1.31 & 1.33 & 1.32 \\
		\midrule
		FasNet+TAC \cite{9054177} & 6.85 & 5.94 & 6.42 & 1.75 & 1.70 & 1.73
		\\
		\midrule
		MIMO-DBnet & \textbf{8.36} & \textbf{8.40} & \textbf{8.38} & \textbf{2.28} & \textbf{2.30} & \textbf{2.29} \\
		\quad \quad w/o DE & 7.81 & 7.76 & 7.78 & 2.18 & 2.15 & 2.16  \\
		\quad \quad w/o DE\&IPD & 6.72 & 6.63 & 6.67 & 2.02 & 1.97 & 2.00 \\

		\bottomrule
        \end{tabular}}
\vspace{-10px}
\label{tbl:overall}
\end{table}
\vspace{-10px}
\begin{table}[t]
\centering
\fontsize{9}{8}\selectfont
\caption{SI-SDR (dB) and PESQ for different frequency subbands on test set. ``$<$ 8 kHz'' and ``$>$ 8 kHz" represent frequency band below 8 kHz and frequency band above 8 kHz, respectively.}
\vspace{5px}
    \setlength{\tabcolsep}{0.5mm}{
        \begin{tabular}{l c c c c c c c}
        \toprule
        \multirow{2}{*}{\textbf{Systems}\ \textbf{/\ Metrics}} & \multicolumn{2}{c}{\textbf{SI-SDR $\uparrow$}} & \multicolumn{2}{c}{\textbf{PESQ $\uparrow$}} \\
        \cmidrule(l){2-3} \cmidrule(l){4-5}
        & $<$ 8 kHz & $>$ 8 kHz
        & $<$ 8 kHz & $>$ 8 kHz\\
		\midrule
		Mixture & -0.22 & -0.30 & 
		1.25 & 1.25 \\
        \midrule
		FaSNet+TAC \cite{9054177} & 5.96 & 5.86 & 1.61 & 1.61  \\
		\midrule
		MIMO-DBnet & \textbf{7.65} & \textbf{7.48} & \textbf{1.98} & \textbf{1.98} \\
		\quad \quad w/o DE & 7.20 & 7.00 & 1.90 & 1.90 \\
		\quad \quad w/o DE\&IPD & 6.24 & 5.99 & 1.79 & 1.79 \\
		\bottomrule
        \end{tabular}}
\vspace{-10px}
\label{tbl:subbband}
\end{table}
\vspace{-1px}
\begin{table}[t]
\centering
\fontsize{9}{8}\selectfont
\caption{Word Error Rate (\%) results in a comparative study of different systems.}
\vspace{5px}
    \setlength{\tabcolsep}{1mm}{
        \begin{tabular}{l c c c c c}
        \toprule
        \multirow{2}{*}{\textbf{Systems}\ \textbf{/\ Metrics}} & \multicolumn{5}{c}{\textbf{WER (\%) $\downarrow$}} \\
        \cmidrule(l){2-6} 
        & $<$ 8 kHz & $>$ 8 kHz & Diff. & Same & Avg.
        \\
		\midrule
		FasNet+TAC \cite{9054177} & 49.40 & 50.93 & 43.37 & 49.70 & 47.70 \\
		\midrule
		MIMO-DBnet & \textbf{13.15} & \textbf{12.90} & \textbf{11.56} & \textbf{11.16} & \textbf{12.34} \\
		\quad \quad w/o DE & 15.49 & 16.72 & 13.93 & 16.56 & 15.86 \\
		\quad \quad w/o DE\&IPD & 24.87 & 24.97 & 20.80 & 21.79 & 23.19  \\
		\bottomrule
        \end{tabular}}
\vspace{-10px}
\label{tbl:wer}
\end{table}
\subsubsection{Analysis of different frequency subbands}
As shown in Table \ref{tbl:subbband}, SI-SDR scores on high frequency band are lower than those on low frequency band. This is due to the \textit{phase wrapping} issue on high frequencies, which has been illustrated in previous sections. In terms of SI-SDR, PESQ and WER, MIMO-DBnet performs constantly better than baseline systems on both high and low frequency subbands. Comparing SI-SDR and WER improvements between MIMO-DBnet and MIMO-DB w/o DE on high and low frequency bands (7.48 - 7.00 dB vs. 7.65 - 7.20 dB, 16.72 - 12.90\% WER vs. 15.49 - 13.15\% WER), we can conclude that MIMO-DBnet is more helpful for recovering high frequencies thanks to the enriched direction discrimination provided by the DOA estimator. Note that high frequencies themselves are good for PESQ, which is why their PESQ scores are no worse than those on low frequencies.
\vspace{-5px}
\subsubsection{Ablation study}
\textbf{MIMO-DBnet w/o DE.} With the assistance of explicit spatial discrimination contained in the DOA-based directional embedding, MIMO-DBnet gained 0.6 dB SI-SDR, 0.13 PESQ and 3.52\% WER absolute improvements compared to MIMO-DBnet w/o DE. This verifies the effect of directional embedding and its beneficial guidance to the neural beamformer for task of speech separation. And the consistent SI-SDR, PESQ and WER results of MIMO-DBnet on different and same gender mixtures also validate the improved spatial discrimination brought by our proposed DOA estimator.

\begin{table}[t]
\centering
\small
\caption{Accuracy (\%) and MAE ($^{\circ}$) results reflecting the frame-level DOA estimation performance of our proposed MIMO-DBnet.}
\vspace{5px}
    \setlength{\tabcolsep}{1mm}{
        \begin{tabular}{c c c c c c c}
        \toprule
        \multirow{2}{*}{\textbf{Metrics}\ \textbf{/\ System}} & \multicolumn{5}{c}{\textbf{MIMO-DBnet}}  \\
        \cmidrule(l){2-6} 
        & $<15^{\circ}$ & $15-45^{\circ}$ & $45-90^{\circ}$ & $>90^{\circ}$ & Avg.
        \\
		\midrule
		Accuracy (\%) $\uparrow$ & 86.22 & 91.91 & 91.40 & 87.16 & 90.34 \\
		\midrule
		MAE ($^{\circ}$) $\downarrow$ & 2.69 & 2.36 & 2.67 & 2.91 & 2.62 \\
		\bottomrule
        \end{tabular}}
\vspace{-10px}
\label{tbl:doa}
\end{table}

As shown in Table \ref{tbl:doa}, we group the DOA results by angular distance between two speakers into four categories, i.e., $<15^{\circ}$, $15-45^{\circ}$, $45-90^{\circ}$ and $>90^{\circ}$.
Considering that 5 degrees is an admissible error \cite{perotin2019crnn}, the DOA estimation performance of MIMO-DBnet is quite outstanding (average angular error of $2.6^{\circ}$), even on closely located cases. This confirms that the improved separation performance of MIMO-DBnet over MIMO-DBnet w/o DE comes from the precisely estimated directional embeddings. The more exactly directional embeddings are predicted, the more completely individual sources are segregated attributed to their guidance. And vice versa, better separation derives from more accurately estimated speech and interference covariance matrices, which also leads to more precise estimation of directional embeddings.

\textbf{MIMO-DBnet w/o DE\&IPD.} Despite MIMO-DBnet w/o DE\&IPD bridges the gap between speech quality on high frequencies and low frequencies to a certain degree (24.97\% vs. 24.87\% WER) by waiving all the spatial discrimination, its performance on all frequency bands are limited due to the lack of spatial information.

\vspace{-5px}
\section{Conclusion}
\label{sec:conlcusion}
In this work, we design a novel multi-channel input and multiple outputs DOA-aware beamforming network, aiming at providing an alternative feature with spatial discrimination when phase differences are ambiguous on high frequencies due to \textit{spatial aliasing} phenomenon. By introducing a light-weight DOA estimator (1.2M parameters) into neural beamformer, the estimated directional embeddings substitute for the ineffective inter-channel features to offer spatial discrimination when \textit{phase wrapping} occurs on high frequencies.   
Experimental results show that our proposed system performs stunningly on both speech separation and DOA estimation tasks.

\vfill\pagebreak

\bibliographystyle{IEEEbib}
\bibliography{strings,refs}

\end{document}